\documentclass[aps,prl,twocolumn,longbibliography,superscriptaddress]{revtex4-2}
\usepackage{bbm}
\usepackage{graphicx}% Include figure files
\usepackage{dcolumn}% Align table columns on decimal point
\usepackage{bm}% bold math
\usepackage{subfigure}
\usepackage{amsmath}
\usepackage{feynmf}
\usepackage{hyperref}
\usepackage{CJK}
\usepackage{amssymb}
\usepackage{attachfile}
\newcommand{\bk}{\boldsymbol k}

\newcommand{\bq}{\boldsymbol q}

\newcommand{\bM}{\boldsymbol M}

\newcommand{\bA}{\boldsymbol{A}}

\newcommand{\bn}{\boldsymbol{n}}
\newcommand{\bs}{\boldsymbol{s}}
\newcommand{\bh}{\boldsymbol{h}}
\newcommand{\zb}{\color {black}}

\usepackage{braket}
\usepackage{esint}
\usepackage{times}

\begin{document}

\title{Light-induced odd-parity altermagnets on dimerized lattices}

\author{Dongling Liu}
\altaffiliation{These authors contributed equally to this work.}
\affiliation{Guangdong Provincial Key Laboratory of Magnetoelectric Physics and Devices,
State Key Laboratory of Optoelectronic Materials and Technologies,
School of Physics, Sun Yat-sen University, Guangzhou 510275, China}

\author{Zheng-Yang Zhuang}
\altaffiliation{These authors contributed equally to this work.}
\affiliation{Guangdong Provincial Key Laboratory of Magnetoelectric Physics and Devices,
State Key Laboratory of Optoelectronic Materials and Technologies,
School of Physics, Sun Yat-sen University, Guangzhou 510275, China}

\author{Di Zhu}
\altaffiliation{These authors contributed equally to this work.}
\affiliation{Guangdong Provincial Key Laboratory of Magnetoelectric Physics and Devices,
State Key Laboratory of Optoelectronic Materials and Technologies,
School of Physics, Sun Yat-sen University, Guangzhou 510275, China}

\author{Zhigang Wu}
\email{ wuzhigang@quantumsc.cn}
\affiliation{Quantum Science Center of Guangdong-Hong Kong-Macao Greater Bay Area (Guangdong), Shenzhen 508045, China}

\author{Zhongbo Yan}
\email{yanzhb5@mail.sysu.edu.cn}
\affiliation{Guangdong Provincial Key Laboratory of Magnetoelectric Physics and Devices,
State Key Laboratory of Optoelectronic Materials and Technologies,
School of Physics, Sun Yat-sen University, Guangzhou 510275, China}

\date{\today}

\begin{abstract}
Altermagnets are an emerging class of collinear magnets with momentum-dependent
spin splitting and zero net magnetization. These materials can be broadly classified 
into two categories based on the behavior of spin splitting at time-reversal-related 
momenta: even-parity and odd-parity altermagnets. 
While even-parity altermagnets have been thoroughly investigated 
both theoretically and experimentally, the systems capable of hosting odd-parity altermagnetism remain largely unexplored.
In this work, we demonstrate that circularly polarized light dynamically converts 
collinear $\mathcal{PT}$-symmetric antiferromagnets on dimerized lattices into odd-parity $p$-wave altermagnets. 
Because of the underlying Dirac band structure of the dimerized lattice, we find that 
the resulting $p$-wave altermagnets can realize Chern insulators (2D) and Weyl semimetals (3D) 
under appropriate drive conditions. Our findings demonstrate that collinear antiferromagnets 
on dimerized lattices provide ideal platforms to investigate the dynamical generation of 
odd-parity altermagnetism. 
\end{abstract}

\maketitle

Altermagnets (AMs), a recently discovered class of collinear compensated magnets, 
have attracted significant attention for combining the advantageous features of ferromagnets 
and antiferromagnets~\cite{Libor2022AMa,Libor2022AMb,Libor2020AM,Hayami2019AM,Hayami2020AM,Yuan2020AM,
Yuan2021AM,Mazin2021,Shao2021NC,Ma2021AM,Liu2022AM,Chen2023AM,Xiao2023AM,Jiang2023AM,Osumi2024MnTe,Lee2024MnTe,Krempasky2024,
Hajlaoui2024AM,Reimers2024,Ding2024CrSb,Yang2024CrSb,
Zeng2024CrSb,Li2024CrSb,Jiang2024KV2Se2O}. 
Their defining characteristic is momentum-dependent spin splitting (MDSS) 
in the band structure despite  zero net magnetization and the absence of spin-orbit coupling (SOC), 
necessitating the existence of nodal planes across which the spin splitting reverses
sign. Based on the rotation angle between adjacent nodal planes, these systems 
are classified as $d$-, $g$-, and $i$-wave AMs~\cite{Libor2022AMa}. {\zb All of them share a unified electronic structure: 
$E(\bk,\bs)\neq E(\bk,-\bs)$ for generic momenta $\bk$ while maintaining
$E(\bk,\bs)=E(-\bk,\bs)$ ($\bs$ denotes spin). }
This symmetry classifies them as even-parity AMs since the MDSS is even under inversion ($\mathcal{P}$).
Such even-parity MDSS enables numerous extraordinary phenomena, such as 
anisotropic tunneling phenomena\cite{Ouassou2023AM,Sun2023AM,Papaj2023AM,Beenakker2023AM,Cheng2024AM,Nagae2025,Lu2024AM11,Zhao2025AM5}, finite-momentum Cooper pairing~\cite{Zhang2024AM,Sumita2023FFLO,Chakraborty2023AM,Sim2025AM7,Liu2025AM8},
unconventional superconductivity\cite{Brekke2023AM,Kristian2024AM,Bose2024AM11,Carvalho2024AM12,Parthenios2025,Maeda2025AM4,Hong2025AM4,
Parshukov2025,Wu2025AM,Fukaya2025AM8,Fu2025AM5,Fu2025AM7,Yokoyama2025Floquet,Zhu2023TSC}, 
diverse topological phases~\cite{Zhu2023TSC,Zhu2024dislocation,Li2023AMHOTSC,Li2024AMHOTI,Ghorashi2024AM,Rao2024AM7,Ma2024AM8,Antonenko2025AM,Qu2025AM,
Parshukov2025AM7,Fernandes2024AM,Zhuang2025AM2,Liu2025AM7}, {\zb and the formation of 
exotic spin textures when combined with SOC~\cite{Ghorashi2025}}, making them promising candidates for 
next-generation spintronics and quantum computing applications. 

MDSS  in magnets can also be odd under inversion. This demands that the energy bands satisfy the relation:
{\zb $E(\bk,\bs)=E(-\bk,-\bs)$}. The existence of this energy relation in magnets is remarkable 
because it typically arises from time-reversal symmetry (TRS), yet this symmetry 
is explicitly broken in magnets. Although relativistic SOC can generate this effect,  
the pursuit of nonrelativistic mechanisms is of fundamental interest. 
One proposed route involves interaction-driven Fermi surface 
instabilities~\cite{Hirsh1990OPAM,Hirsh1990OPAMb,Gorkov1992,Varma2006,Wu2004soc,Wu2007Fermi},
though its experimental realization remains elusive, potentially for fundamental reasons~\cite{Kiselev2017,Wu2018FS}. 
Alternatively, recent spin group theory has established
that odd-parity MDSS can arise at the single-particle 
level in  coplanar magnets that lack inversion symmetry but preserve 
the combined time-reversal-translation symmetry ($\mathcal{T}\boldsymbol{\tau}$)~\cite{Birk2023}.
This discovery has led to the introduction of the concept of  
$p$-wave magnets~\cite{Birk2023} and their higher-partial-wave generalizations~\cite{Yu2025AM}. 
Remarkably, experimental evidence for odd-parity MDSS in such coplanar magnets 
has recently been reported~\cite{Song2025pwave}, 
demonstrating a viable nonrelativistic mechanism for its realization.

The odd-parity MDSS in these coplanar magnets exhibits two key features~\cite{Birk2023,Yu2025AM,Brekke2024OAM,Xu2025OAM7}:
(1) it is confined to the spin component perpendicular to the moment plane, 
satisfying $\langle \bs_{\perp}(\mathbf{k})\rangle=-\langle \bs_{\perp}(-\mathbf{k})\rangle$; and 
(2) the spin polarization, being non-conserved, displays strong momentum dependence. 
Thus, it requires precise Fermi energy or optical frequency tuning to access the regions of significant 
polarization, severely limiting potential applications. A more favorable scenario would 
involve realizing odd-parity MDSS in collinear magnets.
Remarkably, this is possible.  While earlier spin group theory has only identified even-parity AMs, 
this conclusion originated from the assumption that the underlying nonmagnetic crystal structures preserve TRS. 
Recently, Lin revealed that if TRS is 
preemptively broken by sublattice currents~\cite{Haldane1988}, the subsequent emergence of collinear antiferromagnetic order 
in a bipartite lattice can realize odd-parity MDSS~\cite{Lin2025OAM}. This discovery defines 
odd-parity $p$- and $f$-wave AMs, 
 distinct from even-parity and noncollinear types.

Circularly polarized light (CPL) offers a direct and 
versatile alternative to break TRS. Furthermore, when applied to systems with 
a Dirac band structure, CPL can dynamically induce topologically nontrivial phases, 
such as 2D Chern insulators~\cite{Oka2009,Kitagawa2011Floquet,Gu2011,cayssol2013floquet,Usaj2014} and 3D Weyl semimetals~\cite{wang2014floquet,Hubener2017,yan2016tunable,Chan2016type,
Narayan2016,Yan2017Weyl,Ezawa2017FWSM}. Therefore, irradiating collinear 
antiferromagnets with CPL could serve as a generic approach for engineering odd-parity 
AMs with nontrivial topological properties, a mechanism recently demonstrated  
in 2D honeycomb-lattice systems by three independent groups~\cite{Huang2025AM7,Li2025AM7,Zhu2025AM8}. 
Here, we implement this approach in 2D and 3D collinear antiferromagnets on dimerized lattices, relevant to  strongly correlated materials 
like cuprates~\cite{Cai2021SSHH,Tanjaroon2025}. We demonstrate that CPL drives the system into an odd-parity $p$-wave 
AM. The resulting MDSS is tunable via the handedness and incident direction 
of the CPL and emerges consistently along axes (2D) or in planes (3D)  perpendicular to the dimerization. 
Above a threshold intensity, the system transitions into an altermagnetic Chern insulator (2D) 
or altermagnetic Weyl semimetal (3D).

\begin{figure}[t]
\centering
\includegraphics[width=0.45\textwidth]{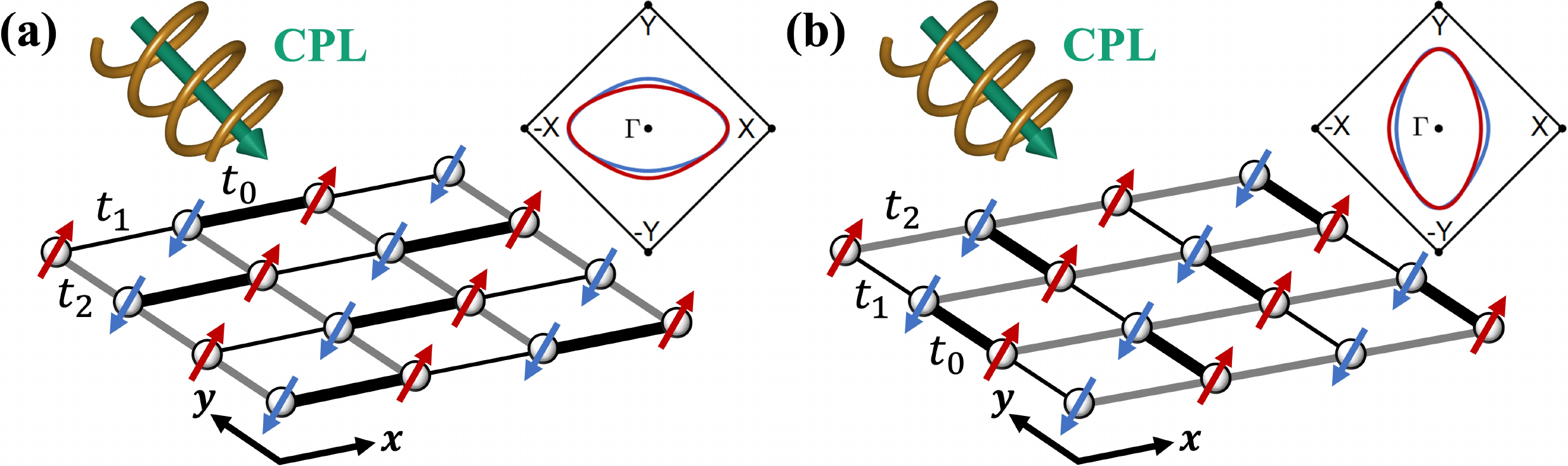}
\caption{Schematic of a collinear antiferromagnet on a 2D dimerized lattice under CPL irradiation. 
Red and blue arrows on the lattice represent magnetic moments of opposite directions.   
Thick (black) solid lines 
denote the bonds of dimerization. (a) 
When dimerization occurs along the $x$ direction (type-I), the Fermi surfaces with opposite spins undergo opposite shifts  along the $y$ direction
{\zb in the presence of CPL}. 
(b) When dimerization is oriented along the $y$ direction (type-II), they instead shift oppositely 
 along the $x$ direction. 
}\label{fig1}
\end{figure}

{\it Two-dimensional collinear antiferromagnets on dimerized lattices.---}We examine two dimerization configurations 
related by a $\pi/2$ rotation, aiming to show that the induced odd-parity MDSS
exhibits a nontrivial dependence on the dimerization direction, as illustrated in Fig.~\ref{fig1}. 
The single-particle Hamiltonian describing spin-$\frac{1}{2}$
fermions propagating on these dimerized lattices under the magnetic background is given by 
\begin{eqnarray}
H=\sum_{\langle ij\rangle}t_{ij}c_{i}^{\dag}c_{j}+
\sum_{i}\xi_{i} c_{i}^{\dag}(\bM\cdot\bs) c_{i},
\end{eqnarray} 
where $c_{i}=(c_{i,\uparrow},c_{i,\downarrow})^{T}$ with $c_{i,s}$ ($c_{i,s}^{\dag}$) 
denoting the annihilation (creation) operator for a fermion at site $i$ and with spin 
$s$. The parameter $t_{ij}$ denotes the hopping constant between two nearest-neighbor sites $i$ and $j$, 
with the real-space patterns illustrated in Fig.~\ref{fig1}. 
The exchange field induced by antiferromagnetic order is given by the vector 
$\bM=(M_{x},M_{y},M_{z})$, while $\bs=(s_{x},s_{y},s_{z})$ denotes the 
vector of Pauli matrices in spin space. The sublattice degree of freedom is encoded by $\xi=1$ (sublattice A) 
and $\xi=-1$ (sublattice B), reflecting the staggered nature of the antiferromagnetic order.

By performing a Fourier transformation, the corresponding momentum-space Hamiltonian 
for the type-I dimerization configuration shown in Fig.~\ref{fig1}(a) is obtained as 
\begin{eqnarray}
\mathcal{H}^{\rm (I)}(\bk)&=&(t_{+}\cos k_{x}+2t_{2}\cos k_{y})\sigma_{x}+t_{-}\sin k_{x}\sigma_{y}\nonumber\\
&&+\bM\cdot\bs \sigma_{z},\label{typeI}
\end{eqnarray} 
where the Pauli matrices $\sigma_{i}$ act on the two sublattice degrees of freedom, 
and the parameters $t_{\pm}=t_{0}\pm t_{1}$. For notational simplicity, 
we set the lattice constants to unity, omit identity matrices, and assume 
all three hopping parameters ($t_{0}$, $t_{1}$ and $t_{2}$) to be positive constants throughout this work. 
The Hamiltonian for the type-II dimerization configuration can be obtained from the type-I Hamiltonian through the relation $\mathcal{H}^{\rm (II)}(k_{x},k_{y})=\mathcal{H}^{\rm (I)}(k_{y},-k_{x})$. 
Given this straightforward mapping, we will focus our analysis on $\mathcal{H}^{\rm (I)}(\bk)$ 
in the following discussion. 

In the absence of antiferromagnetic order ($\bM=0$), the Hamiltonian 
describes AB-stacked one-dimensional Su-Schrieffer-Heeger (SSH) chains~\cite{Su1979} along the $y$ direction. 
In the weak-coupling regime ($t_{2}\ll t_{+}$), the system forms a weak topological 
insulator exhibiting flat-band edge states exclusively on $x$-normal edges. 
These flat bands float within the bulk gap and span the entire boundary Brillouin zone. 
Conversely, in the strong-coupling regime ($t_{2}\gg t_{-}$), the system becomes 
a Dirac semimetal that also supports flat bands on $x$-normal edges. However,  
these flat bands terminate at the projections of bulk Dirac points and thus only 
occupy a portion of the boundary Brillouin zone.  In what follows, we 
focus on the regime with $t_{2}>{\zb t_{+}}/2$, corresponding to a Dirac 
semimetal when $\bM=0$.

{\zb For the Hamiltonian in Eq.~(\ref{typeI}), the exchange field breaks 
both inversion symmetry ($\mathcal{P}=\sigma_{x}$) and spinful TRS ($\mathcal{T}_{1}=is_{y}\mathcal{K}$).
 Nevertheless, the combined 
$\mathcal{P}\mathcal{T}_{1}$ symmetry remains intact. Since this combined 
operation squares to $-1$ and commutes with the Hamiltonian, Kramers theorem 
guarantees that all bands are spin-degenerate. Interestingly, when the exchange field 
is absent ($\bM=0$), the Hamiltonian becomes spin-independent and acquires an 
emergent spinless $\mathcal{P}\mathcal{T}_{2}$ symmetry, where 
$\mathcal{T}_{2}=\mathcal{K}$. Here, we introduce a subscript to distinguish these two 
distinct TRS operators.  
With $(\mathcal{P}\mathcal{T}_{2})^{2}=1$, 
this symmetry protects twofold Dirac points in each spin sector by ensuring 
a quantized $\pi$ Berry phase around them. A nonzero exchange field ($|\bM|\neq0$) 
breaks this symmetry, gapping the Dirac points and 
driving the system into an insulating phase (see Sec.~I of the Supplemental Material (SM)~\cite{supplemental} for details). }

{\it Floquet $p$-wave altermagnetic Chern insulator.---}Now we explore the influence of 
CPL on the system. Consider light incident perpendicular to the plane of the 2D system, 
described by the vector potential 
$\bA(t)=A_{0}(\cos\omega t,\eta \sin\omega t)$, where $\eta=+1$ and $\eta=-1$ correspond to 
right-handed and left-handed CPL, respectively. The electromagnetic coupling is given by 
 $\mathcal{H}^{\rm (I)}(\bk)\rightarrow \mathcal{H}^{\rm (I)}[\bk+\bA(t)]$ (we set $e=\hbar=1$ for notational simplicity).
Since the Hamiltonian becomes time-periodic, it can be expanded in a Fourier series as 
$\mathcal{H}^{\rm (I)}[\bk+\bA(t)]=\sum_{n}\mathcal{H}_{n}(\bk)e^{i n\omega t}$ with  $n\in \mathbb{Z}$.  
{\zb The explicit expressions of $\mathcal{H}_{n}$ for $|n|\leq2$ are provided in Sec.~II of the SM~\cite{supplemental}}. 
We focus on the high-frequency off-resonant regime ({\zb the frequency is beyond the bandwidth}), 
where the driven system is effectively described by a time-independent Hamiltonian 
given by~\cite{Kitagawa2011Floquet,Goldman2014}  
\begin{eqnarray}
\mathcal{H}_{\rm eff}(\bk)&=&\mathcal{H}_{0}(\bk)+\sum_{n\geq1}\frac{[\mathcal{H}_{+n},\mathcal{H}_{-n}]}{n\omega}+O(\omega^{-2}).
\end{eqnarray}
{\zb Keeping the leading-order contributions from $\mathcal{H}_{0}$ and $\mathcal{H}_{\pm1}$ (the justification of this simplification 
is analyzed in Sec.~II of the SM~\cite{supplemental})}, we find 
\begin{eqnarray}
\mathcal{H}_{\rm eff}(\bk)&=&J_{0}(A_{0})[({\zb t_{+}}\cos k_{x}+2t_{2}\cos k_{y})\sigma_{x}+{\zb t_{-}}\sin k_{x}\sigma_{y}]\nonumber\\
&&+\bM\cdot\bs\sigma_{z}-\eta F(A_{0},\omega)\sin k_{y}\cos k_{x}\sigma_{z},\label{FloquetH}
\end{eqnarray}
where $J_{n}(x)$ is the $n$-th order Bessel function of the first kind,  
and $F(A_{0},\omega)=8J^2_{1}(A_{0}) t_{2}{\zb t_{-}}/\omega$. Compared to 
the original static Hamiltonian, the key modification induced by CPL is the emergence of the last term. 
This term breaks the spinful $\mathcal{P}\mathcal{T}_{1}$ symmetry of the static Hamiltonian, thereby lifting the spin degeneracy 
in the band structure. Notably, this term exhibits odd-parity momentum dependence, which is essential 
for generating odd-parity MDSS. We emphasize that this term arises from the interplay 
between the $\cos k_{y}\sigma_{x}$ and $\sin k_{x}\sigma_{y}$ terms in the time-periodic Hamiltonian.
Three key features of these two terms---their anticommutation relation, dependence on orthogonal momenta, 
and opposite parity under momentum inversion (even vs odd)---are responsible for producing terms with the desired 
odd-parity momentum dependence. 
Importantly, terms fulfilling these three conditions appear generically in dimerized systems, 
suggesting the universality of this mechanism. 
  
Although both TRS and spinful $\mathcal{P}\mathcal{T}$ symmetry are broken, the driven system maintains compensated magnetization. 
This compensation arises from the preserved $[C_{2\perp}\|\mathcal{P}]$ symmetry in the driven Hamiltonian, where
$C_{2\perp}$ denotes a $\pi$ rotation in the spin space around an axis perpendicular to the Néel vector, 
and $\mathcal{P}$ denotes the spatial inversion operation. This symmetry guarantees the energy relation
$E(\bk,\bs)=E(-\bk,-\bs)$, which explicitly forbids net magnetization. Notably, 
$[C_{2\perp}\|\mathcal{P}]$  imposes the same constraint on energy bands as TRS. 
Consequently, when spin splitting occurs, it must exhibit an odd-parity pattern---analogous 
to that induced by SOC.

\begin{figure}[t]
\centering
\includegraphics[width=0.45\textwidth]{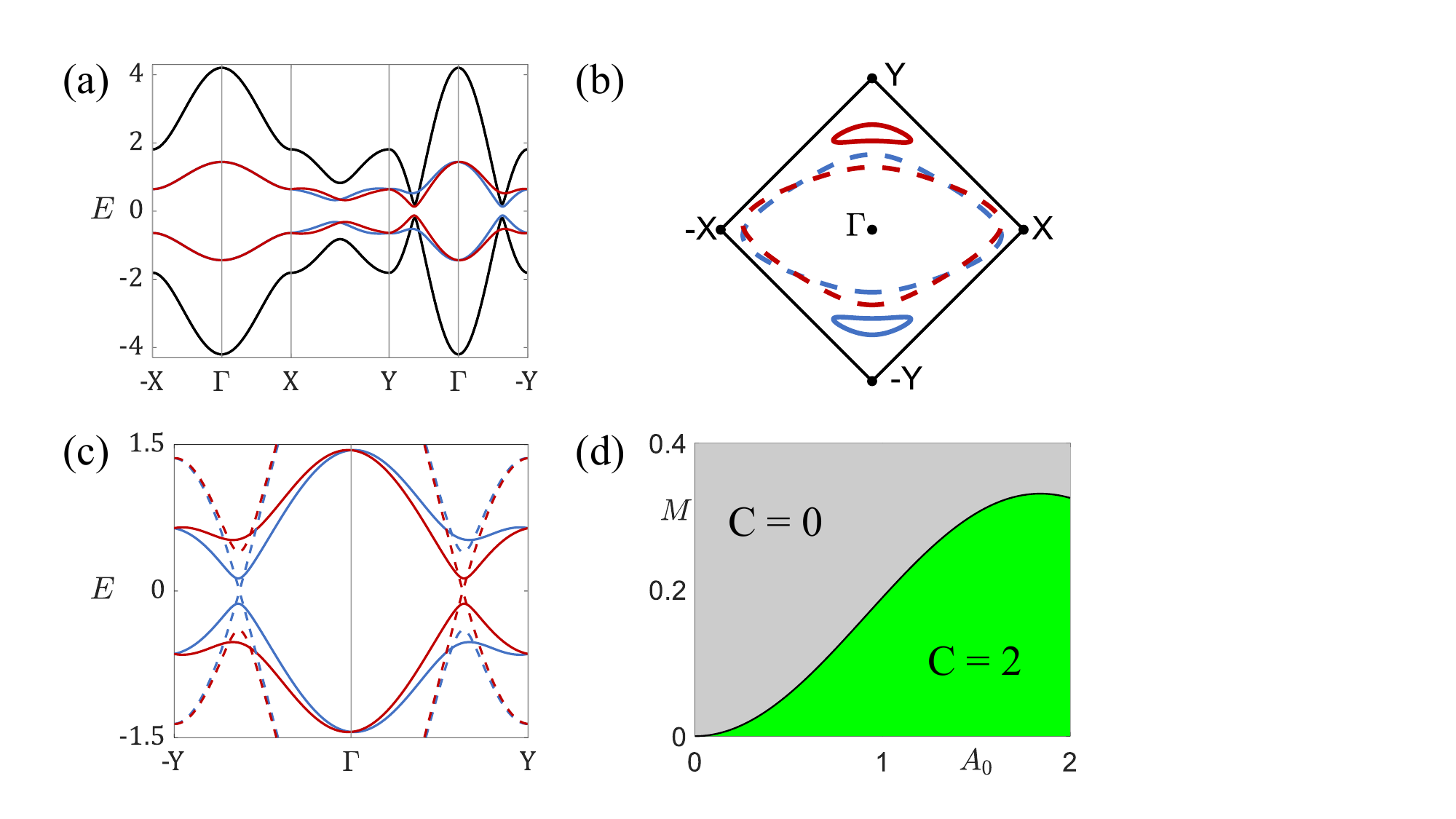}
\caption{(a) Energy bands of the static (black, spin-degenerate) 
and CPL-driven ($A_0 = 1.8$, red/blue for spin up/down) systems.  
(b)  Fermi surface at Fermi energy $E_{F}=0.2$ (solid lines) and 
$E_{F}=0.7$ (dashed lines), corresponding to the light-induced spin-split bands in (a). 
(c) Band gap evolution: closure at the critical drive (dashed, $A_0 = 1.04$) and reopening for a stronger drive (solid, $A_0 = 1.8$).
(d) Topological phase diagram in the $(M, A_0)$ plane. {\zb The phase boundary separating 
the two regions with $C=0$ and $C=2$ is determined by the gap-closing condition: $M=\tilde{F}$}. The value $M = 0.2$ is used in (a-c). 
Shared parameters for all panels are $t_{0}=1$, $t_{1}=0.2$, $t_{2}=1.5$, $\eta=1$, and $\omega=9$.  
}\label{fig2}
\end{figure}

Given the system's rotational invariance with respect to the Néel vector orientation, 
we hereafter fix the Néel vector along the $z$ direction, i.e., $\bM=M\hat{z}$ with $M>0$.
In Fig.~\ref{fig2}(a), we show a comparison of the band structure with and without CPL irradiation.
In the static case, the spin-up and spin-down energy bands are degenerate. Under CPL driving, 
however, a clear spin splitting emerges in the band structure. This splitting exhibits two key features:  odd-parity nature and strong 
anisotropy. The odd-parity feature is evident in the energy spectrum along the $k_{y}$ axis (-$\text{Y}$-$\Gamma$-$\text{Y}$). 
The anisotropy is manifested as a stark contrast in the energy spectrum between  $k_{x}$ (-$\text{X}$-$\Gamma$-$\text{X}$) and $k_{y}$ axes:
the splitting is strongest along $k_{y}$ but entirely absent along $k_{x}$. 
In Fig. \ref{fig2}(b), we present Fermi surfaces at two distinct Fermi energies ($E_{F}$). 
The spin-up and spin-down Fermi surfaces maintain equal sizes and 
are mutually related through the $[C_{2\perp}\|\mathcal{P}]$ symmetry, further confirming 
the odd-parity character of the spin splitting. {\zb The Fermi surface possesses only a $C_{2}$ 
rotational symmetry. Based on the sign of the spin splitting under a $\pi$ rotation, we can 
classify the splitting into two distinct channels: the $l=0$ (even-parity $s$-wave) and 
the $l=1$ (odd-parity $p$-wave) channels. The observed splitting corresponds to the $l=1$ channel, 
identifying the resulting phase as a $p$-wave AM. To realize higher-partial-wave odd-parity 
AMs, such as an $f$-wave ($l=3$) state, the system must possess correspondingly higher rotational symmetry~\cite{Huang2025AM7,Li2025AM7,Zhu2025AM8}.}

Assuming that the CPL irradiation does not modify the antiferromagnetic order, 
we note that increasing the light amplitude can induce a closing and 
reopening of the bulk energy gap, as illustrated in Fig.~\ref{fig2}(c). 
This gap closure triggers a topological phase transition 
from a trivial insulator to a Chern insulator characterized by 
Chern number $|C|=2$. To intuitively understand this transition, 
we perform a low-energy expansion near 
the momentum where the gap closure occurs. Such momenta 
are at $\bk_{\rm D,\pm}=\pm(0,\pi-\arccos(t_+/2t_{2}))$ (we 
assumed $t_{2}>t_+/2$). Near these two momenta, 
the low-energy Hamiltonian is given by 
\begin{eqnarray}
\mathcal{H}_{\chi,s}(\bq)=-\chi v_{y}q_{y}\sigma_{x}+v_{x}q_{x}\sigma_{y}
+(sM-\chi\eta \tilde{F})\sigma_{z}.
\end{eqnarray}
Here, $\chi=\pm1$ indexes the two valleys located at $\bk_{\rm D,\pm}$, 
while $s = \pm 1$ corresponds to the spin-up and spin-down sectors, respectively. 
The small momentum $\mathbf{q} = (q_x, q_y)$ is measured relative 
to $\mathbf{k}_{\rm D,\pm}$. For notation convenience, we introduce the following definitions:
$v_{x}=J_{0}(A_{0})t_-$, $v_{y}=J_{0}(A_{0})\sqrt{4t_{2}^{2}-t_+^{2}}$, and 
$\tilde{F}=F(A_{0},\omega)\sqrt{4t_{2}^{2}-t_+^{2}}/2t_{2}$.
 
Since the spinful $\mathcal{P}\mathcal{T}$ symmetry guarantees that the pre-driving Hamiltonian 
has $C=0$, the system remains topologically trivial before the 
energy gap closes. When the gap reopens, the Chern number of the driven 
system can be determined by tracking its change through the gap-closing 
transition, as captured by the low-energy Hamiltonians. The Chern numbers 
for those low-energy Hamiltonians are given by~\cite{qi2005} 
\begin{eqnarray}
C_{\chi,s}=-\frac{1}{4\pi}\int dq_{x}\int dq_{y}\frac{\bh_{\chi,s}\cdot(\partial_{q_{x}}\bh_{\chi,s}\times 
\partial_{q_{y}}\bh_{\chi,s})}{|\bh_{\chi,s}|^{3}}, 
\end{eqnarray}
where $\bh_{\chi,s}=(-\chi v_{y}q_{y},v_{x}q_{x},sM-\chi\eta \tilde{F})$. Direct calculation 
yields: $C_{\chi,s}=\text{sgn}(\eta \tilde{F}-\chi sM)/2$, 
where $\text{sgn}(x)$ is the sign function ($+1$ for $x>0$ and $-1$ for $x<0$). Two of these Chern numbers 
undergo a sudden change when $\tilde{F}$ crosses the critical value $M$ at which the bulk energy gap closes. 
Specifically, the changes in Chern number satisfy  
$\Delta C_{+,\uparrow}=\Delta C_{-,\downarrow}=(1+\eta)/2$
and $\Delta C_{-,\uparrow}=\Delta C_{+,\downarrow}=(\eta-1)/2$, which are also a consequence of the $[C_{2\perp}\|\mathcal{P}]$ symmetry. 
Accordingly, the Chern number characterizing the system in the region $\tilde{F} > M$ 
is derived as $C = 2\eta$, which is controlled by the CPL's handedness. 
This analysis fully determines the topological phase diagram shown in Fig. \ref{fig2}(d).

\begin{figure}[t]
\centering
\includegraphics[width=0.45\textwidth]{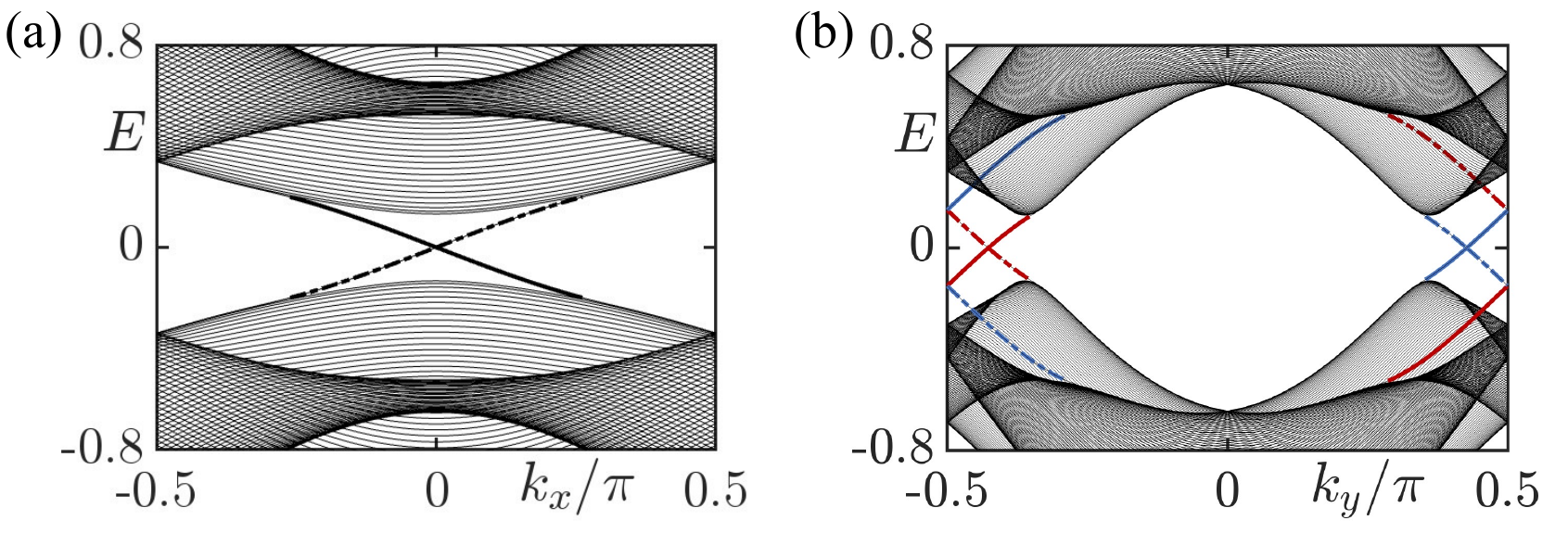}
\caption{Energy spectrum for a sample of cylindrical geometry. 
(a) $y$-normal edges: Spin-degenerate spectrum of chiral edge states. 
States on the bottom (top) edge are shown as solid (dashed) lines.
(b) $x$-normal edges: Spin-split spectrum of chiral edge states. 
States on the left (right) edge are shown as solid (dashed) lines.
Parameters are  $t_{0}=1$, $t_{1}=0.2$, $t_{2}=1.5$, $M=0.2$, $\eta=1$, $A_{0}=1.8$, and $\omega=9$. 
}\label{fig3}
\end{figure}

The emergence of gapless chiral edge states
constitutes a defining signature of Chern insulators. 
Our calculations of the energy spectrum for cylindrical-geometry samples 
demonstrate two branches of gapless chiral edge states in the $\tilde{F} > M$ regime, confirming the predicted bulk topological invariant. 
Strikingly, these edge states inherit the bulk's anisotropic spin splitting: 
while remaining spin-degenerate along $y$-normal edges (where $k_x$ is conserved), 
they develop pronounced spin splitting along $x$-normal edges (with conserved $k_y$), 
as clearly evidenced in Figs.~\ref{fig3}(a) and \ref{fig3}(b).

\begin{figure}[t]
\centering
\includegraphics[width=0.45\textwidth]{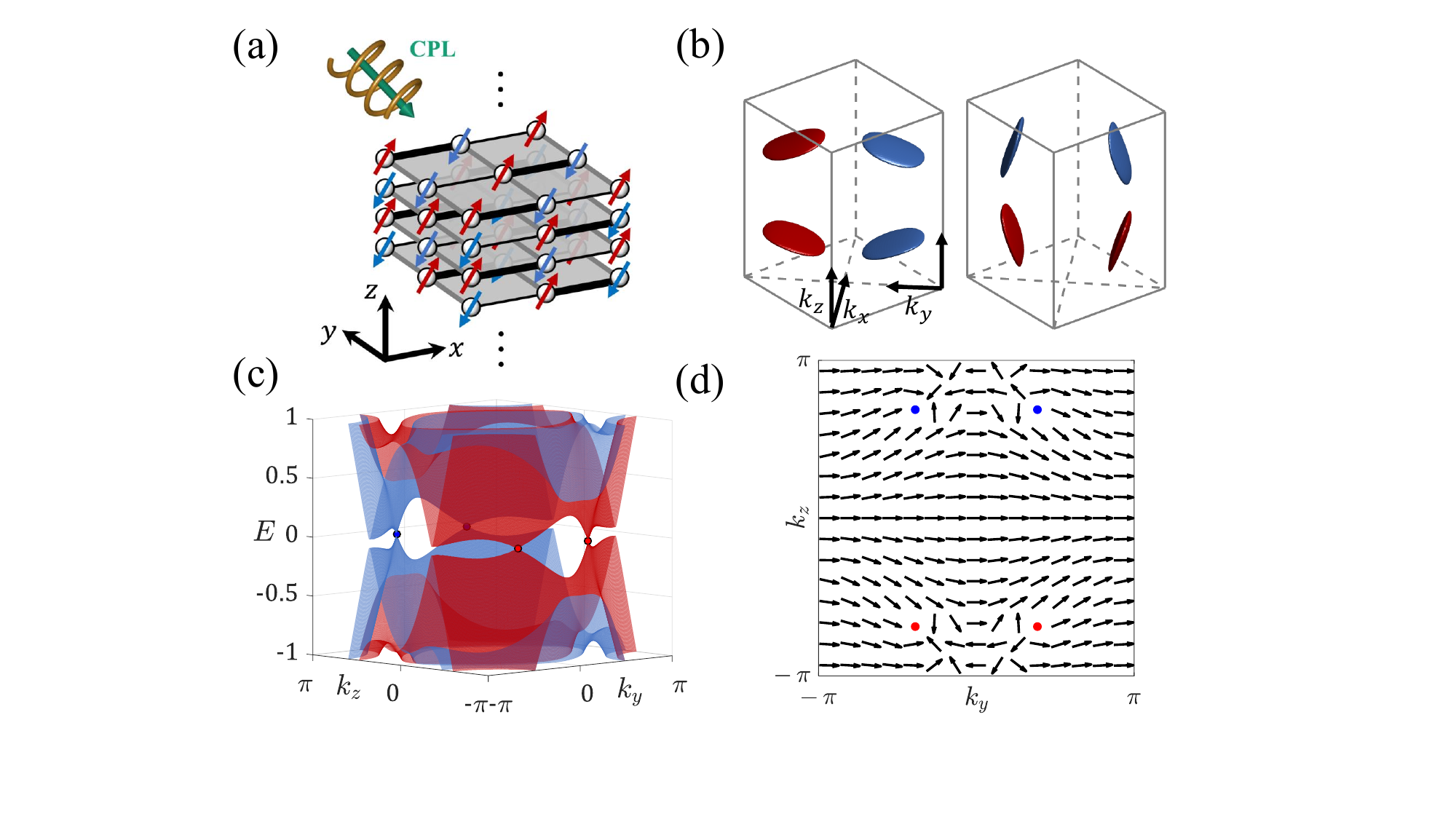}
\caption{(a) Realization of the 3D system through AB stacking. (b) 
{\zb Fermi surfaces of the CPL-driven system at a Fermi energy 
$E_{F}=0.8$. The red and blue contours represent the spin-up and spin-down branches, respectively. Left panel: CPL incident along the 
$z$ direction. Right panel: CPL incident along the $y$ direction}.  
(c) Energy spectrum in the $k_x = 0$ plane, 
showing two Weyl points of each spin.  (d) Berry curvature vector field in the $k_{x}=0$ plane
containing the Weyl points.  
{\zb Arrows show the in-plane components $(\Omega_{y},\Omega_{z})$. 
Red and blue dots mark the Weyl points, which act as Berry flux sources (arrows flowing out) 
and sinks (arrows flowing in), respectively.}
The value $M=1$ is used in (b), and $M=0.4$ is used in (c) and (d). 
Other shared parameters are $t_{0}=1$, $t_{1}=0.2$, 
$t_{2}=0.5$, $t_{3}=2$, $\eta=1$, $A_{0}=1.8$, and $\omega=9$. 
}\label{fig4}
\end{figure}

{\it Floquet $p$-wave altermagnetic Weyl semimetals.---}The 2D scenario 
admits a natural extension to 3D. As a prototypical example, we construct 
a 3D system by AB-stacking identical copies of the 2D Hamiltonian along the $z$ axis [illustrated in 
Fig.~\ref{fig4}(a)], 
obtaining the following static Hamiltonian:
\begin{eqnarray}
\mathcal{H}^{\rm 3D}(\bk)&=&({\zb t_{+}}\cos k_{x}+2t_{2}\cos k_{y}+2t_{3}\cos k_{z})\sigma_{x}\nonumber\\
&&+{\zb t_{-}}\sin k_{x}\sigma_{y}+Ms_{z}\sigma_{z},
\end{eqnarray}
where $t_{3}$ is the interlayer hopping constant. 
Interestingly, we find that  the orientation of incident light has a strong impact on the spin-splitting features
in 3D. To demonstrate this, we consider that the light is incident along a general direction, 
described by $\bn = (\cos\phi \sin \theta,\sin\phi \sin \theta,\cos \theta)$, 
where $\phi$ and $\theta$ are the azimuthal and polar angles on the Bloch sphere, respectively. 
The vector potential of the light is given by 
$\bA(t) = A_{0}[\cos(\omega t)\mathbf{e}_{1} + \eta \sin(\omega t)\mathbf{e}_{2}]$, where 
$\mathbf{e}_{1} = (\sin \phi, -\cos\phi,0)$ and $\mathbf{e}_{2} = (\cos \phi \cos\theta,\sin\phi\cos\theta,-\sin\theta)$, 
corresponding to two unit orthogonal vectors perpendicular to $\bn$~\cite{Liu2025AM7}. 
Following the same derivation as before, we retain the leading-order contributions from $\mathcal{H}_{0}$ and $\mathcal{H}_{\pm1}$ 
(the influence of terms contributed by multiple photon processes, such as $\mathcal{H}_{\pm2}$, 
is examined in Sec.~III of the SM~\cite{supplemental}).
The resulting Floquet Hamiltonian for the driven system is
\begin{eqnarray}
\mathcal{H}^{\rm 3D}_{\rm eff}(\bk)&=&\sum_{\mu=\{x,y,z\}}\tilde{t}_{\mu}\cos k_{\mu}\sigma_{x}+{\zb t_{-}} J_0(A_{1})\sin k_x\sigma_y\nonumber\\
&&+Ms_{z}\sigma_{z}+[F_{y}\sin k_{y}+F_{z}\sin k_{z}]\cos k_{x}\sigma_{z}.\quad\label{3DH}
\end{eqnarray}
For convenience, we have introduced the following shorthand notations: 
$\tilde{t}_{x}={\zb t_+} J_0(A_{1})$, $\tilde{t}_{y}=2t_2 J_0(A_{2})$, 
$\tilde{t}_{z}=2t_3 J_0(A_{3})$, $F_{y}=8J_{1}(A_{1})J_{1}(A_{2}){\zb t_{-}}t_{2}\sin(\phi_{2}-\phi_{1})/\omega$, 
$F_{z}=8J_{1}(A_{1})J_{1}(A_{3}){\zb t_{-}}t_{3}\sin\phi_{1}/\omega$,
where $A_{i}=\lambda_{i}A_{0}$, $\lambda_1=\sqrt{1-\cos^2 \phi \sin^2 \theta}$, 
$\lambda_2=\sqrt{1-\sin^2 \phi \sin^2 \theta}$, $\lambda_3=\eta \sin \theta$,
$\phi_{1}=\arg[\eta\cos \phi\cos\theta+i\sin\phi]$, and $\phi_{2}=\arg[\eta\sin \phi\cos\theta-i\cos\phi]$. 
Despite the added complexity introduced by general-angle incidence, the resulting spin-splitting 
features and band topology remain straightforward to analyze. 

Given the apparent symmetry between $k_{y}$ and $k_{z}$ in the Hamiltonian, Fermi 
surfaces with opposite spins can now be shifted in opposite directions along any axis 
in the $k_{y}$-$k_{z}$ plane. To illustrate this symmetry, we plot the spin-split Fermi 
surfaces under CPL incident along the $y$ and $z$ directions. As shown in Fig. \ref{fig4}(b), 
Fermi surfaces with opposite spins exhibit opposite displacements: along the $z$ direction 
for $y$-propagating CPL ({\zb left panel}), and along the $y$ direction for $z$-propagating CPL ({\zb right panel}). 
{\zb For CPL propagating along a generic direction in the $yz$ plane, defined by
$(\phi,\theta)=(\pi/2,\theta)$ or $\bn=(0,\sin\theta,\cos\theta)$, the parameters 
of the Floquet Hamiltonian are: $\phi_{1}=\pi/2$, $\phi_{2}=[1+\eta\text{sgn}(\cos\theta)]\pi/2$, $\lambda_{1}=1$, 
$\lambda_{2}=|\cos\theta|$, and $\lambda_{3}=\eta\sin\theta$. Consequently, we have
$F_{y}=-8\eta \text{sgn}(\cos\theta)J_{1}(A_{0})J_{1}(A_{0}|\cos\theta|)t_{-}t_{2}/\omega$, and 
$F_{z}=8J_{1}(A_{0})J_{1}(\eta A_{0}\sin\theta) t_{-}t_{3}/\omega$. In the case 
where the $y$ and $z$ directions are equivalent ($t_{2}=t_{3}$), one gets the ratio
$F_{y}/F_{z}=-\eta \text{sgn}(\cos\theta)J_{1}(A_{0}|\cos\theta|)/J_{1}(\eta A_{0}\sin\theta)=-J_{1}(A_{0}\cos\theta)/J_{1}(A_{0}\sin\theta)$.
Now, considering the weak-driving regime ($A_{0}\ll1$) and momenta near $(k_{y},k_{z})=(0,0)$, 
we can use the approximations $J_{1}(x)\approx x/2$ and $\sin k\approx k$, The ratio becomes 
 $F_{y}/F_{z}\approx-\cos\theta/\sin\theta$, and the key term in the Hamiltonian has the property:
 $F_{y}\sin k_{y}+F_{z}\sin k_{z}\propto -\cos\theta k_{y}+\sin\theta k_{z}$. This term vanishes along the direction
 $(k_{y},k_{z})\parallel (\sin\theta,\cos\theta)$ and is maximized along the perpendicular direction 
 $(k_{y},k_{z})\parallel (-\cos\theta,\sin\theta)$. Consequently, Fermi surfaces with opposite spins 
 are displaced oppositely along $(-\cos\theta,\sin\theta)$, which is perpendicular to 
 the light's in-plane propagation direction $(\sin\theta,\cos\theta)$. This analysis 
 reveals that when light propagates perpendicular to the dimerization axis, 
 the three vectors---the dimerization axis, the light's propagation direction, and the 
 Fermi surface displacement---form an approximately orthogonal triad.}

In our 2D system, we demonstrate that a {\zb $p$-wave altermagnetic} Chern insulator  emerges 
when the CPL amplitude surpasses a critical threshold. Stacking 
such Chern insulators can produce either 3D  {\zb $p$-wave altermagnetic} Chern insulators (for weak interlayer coupling) 
or Weyl semimetals (for strong coupling){\zb~\cite{Hasan2010,Qi2011}}. Here, we focus on the realization of {\zb $p$-wave altermagnetic} Weyl semimetals. 
From the 3D Hamiltonian [Eq. (\ref{3DH})], Weyl points emerge 
in the $k_{x}=0$ plane when the CPL amplitude surpasses another critical threshold. 
Their positions are determined by: $\tilde{t}_{x}+\tilde{t}_{y}\cos k_{y}+\tilde{t}_{z}\cos k_{z}=0$
and $s M+F_{y}\sin k_{y}+F_{z}\sin k_{z}=0$ ($s=\pm1$ for spin up/down). 
This system contains a minimum of four Weyl points,
mirroring the situation in noncentrosymmetric Weyl semimetals with TRS~\cite{Armitage2018RMP}. 
Notably, although the unitary $[C_{2\perp}\|\mathcal{P}]$ symmetry constrains 
Weyl point locations identically to antiunitary TRS, it imposes opposite rules 
on the monopole charges of symmetry-related points. Specifically, Weyl points 
related by $[C_{2\perp}\|\mathcal{P}]$ carry opposite charges, whereas those 
related by TRS carry the same charge.

To demonstrate our analysis, we consider the 
simplified case of $y$-direction ($\phi=\theta=\pi/2$) right-handed ($\eta=1$) CPL incidence for clarity.
In this configuration, the possible spin-up Weyl points are located at $\bk_{w1}=(0,k_{y}^{+},k_{z}^{w})$, 
$\bk_{w2}=(0,-k_{y}^{+},k_{z}^{w})$, $\bk_{w3}=(0,k_{y}^{-},\pi-k_{z}^{w})$, $\bk_{w4}=(0,-k_{y}^{-},\pi-k_{z}^{w})$, 
where $k_{z}^{w}=-\arcsin(M/F_{z})$, and $k_{y}^{\pm}=\pi-\arccos[(\tilde{t}_{x}\pm\tilde{t}_{z}\cos k_{w}^{z})/\tilde{t}_{y}]$.
By choosing the parameters such that $|\tilde{t}_{x}-\tilde{t}_{z}\cos k_{w}^{z}|<\tilde{t}_{y}<|\tilde{t}_{x}+\tilde{t}_{z}\cos k_{w}^{z}|$, 
only the two Weyl points at $\bk_{w3}$ and $\bk_{w4}$ are present. Correspondingly, 
there are two spin-down Weyl points located at $-\bk_{w3}$ and $-\bk_{w4}$.
In Fig.~\ref{fig4}(c), we plot the energy bands in the $k_x = 0$ plane, which clearly 
exhibit two Weyl cones in each spin sector. Furthermore, Fig.~\ref{fig4}(d) 
shows the distribution of Berry curvature vector field $\boldsymbol{\Omega}(\bk)=(\Omega_{x}(\bk),\Omega_{y}(\bk),\Omega_{z}(\bk))$, 
where $\Omega_{a}(\bk)=\sum_{E_{n}<0}i\epsilon_{abc}\langle \partial_{k_{b}}u_{n}|\partial_{k_{c}}u_{n}\rangle$. Here
$\epsilon_{abc}$ is the Levi-Civita symbol and 
$|u_{n}(\bk)\rangle$ is the eigenstate of the $n$-th occupied band. {\zb The field clearly shows 
that inversion-related Weyl points carry opposite monopole charges~\cite{Armitage2018RMP}. Specifically, a 
spin-down Weyl point at the top right (left) acts as a source (sink) of Berry flux, while its inversion partner---a spin-up 
Weyl point at the bottom left (right)---acts as a sink (source).}

{\it Discussions and conclusions.---}We demonstrate that CPL can 
dynamically induce odd-parity altermagnetism in collinear 
$\mathcal{PT}$-symmetric antiferromagnets on dimerized lattices. 
The resulting {\zb $p$-wave} altermagnetic phase inherently supports 
topological states due to the underlying Dirac band structure. 
{\zb Experimentally,  the characteristic spin splitting can be detected 
by measuring the spin-split band structure along high symmetry momentum lines 
through spin- and angle-resolved photoemission spectroscopy.}

The antiferromagnets we consider are characterized by the coexistence of bond order and antiferromagnetic order. 
The bond order typically arises from electron-phonon interactions, whereas the antiferromagnetic order 
is driven by strong electron-electron repulsion. Their coexistence has been 
rigorously established through numerical studies of strongly correlated systems~\cite{Tanjaroon2025}. 
Beyond correlated materials, cold-atom systems offer a promising alternative platform for realizing our proposed 
odd-parity AMs. In such systems, the dimerized lattice can be directly 
engineered through optical lattice design~\cite{Atala2013}, while the antiferromagnetic order 
can be stabilized by tuning the on-site interactions to be strongly repulsive~\cite{Hart2015,Mazurenko2017,Shao2024}.
The effect of CPL can be equivalently realized by periodically modulating the lattice~\cite{Jotzu2014}. 

As an outlook, odd-parity altermagnets also host favorable conditions for 
the emergence of unconventional superconductivity, as their band structure mimics 
time-reversal-invariant systems with SOC while 
explicitly breaking TRS. The fundamental relation 
$E(\mathbf{k},\bs) = E(-\mathbf{k},-\bs)$ 
ensures that a spin-up electron always has a spin-down 
partner available for pairing. This opens the door to realizing exotic, 
time-reversal-breaking spin-singlet superconductors.

In conclusion, our work establishes a general dynamical protocol for creating odd-parity 
AMs, significantly expanding the class of systems capable of hosting these exotic phases 
and their emergent phenomena.

{\it Acknowledgements.---}This work is supported by the National Natural Science Foundation of China (Grant No. 12174455, No. 12474264),  Guangdong Basic and Applied Basic Research Foundation (Grant No. 2023B1515040023), Guangdong Provincial Quantum Science Strategic Initiative (Grant No. GDZX2404007) and
National Key R\&D Program of China (Grant No. 2022YFA1404103).

\bibliography{dirac.bib}

\end{document}